\documentclass[11pt]{iopart}
\usepackage{multirow} 
\usepackage[figuresright]{rotating}
\pdfoutput=1

\begin{document}
\title{Performance of a SensL-30035-16P Silicon Photomultiplier array at liquid argon temperature} 


\author{S. Catalanotti$^a$, A. G. Cocco$^a$, G. Covone$^a$, M. D'Incecco$^c$, G. Fiorillo$^a$, G.~Korga$^{d}$, B.~Rossi$^{a,b}$\thanks{Corresponding author.}, S. Walker$^{a}$ }
\address{$^a$Universit\`a degli Studi di Napoli "Federico II" \& INFN Sezione di Napoli, Via Cinthia, I-80125 Napoli, Italy}
\address{$^b$Department of Physics, Princeton University, Princeton, New Jersey 08544, USA}
\address{$^c$INFN, Laboratori Nazionali del Gran Sasso, Via G. Acitelli 22, I-67010 Assergi (AQ), Italy}
\address{$^d$Department of Physics, University of Houston, Houston, TX 77204, USA}

\ead{biagio.rossi@na.infn.it}

\begin{abstract}
Next generation multi-ton scale noble liquid experiments have the unique opportunity to discover dark matter particles at the TeV scale, reaching the sensitivity of 10$^{-48}$ cm$^2$ in the WIMP nucleon scattering cross-section. A prerequisite will be the reduction of radiogenic background sources to negligible levels. This is only possible if ultra-pure high efficiency photosensors are available for the scintillation light readout. Current experiments (e.g. Xenon, LUX, Darkside, ArDM) use cryogenic PMTs as photosensors. An attractive alternative is represented by silicon photomultiplier arrays (SiPM arrays), which show unrivalled performances in single photon detection. This paper reports on the performance of the SensL-ArrayB-30035-16P SiPM array and a custom made cryogenic front-end board at the liquid argon temperature. Its performance at V$_\mathrm{OV}$=3.5~V, where the PDE is maximal, are very promising in terms of SPE resolution (about 8\%), dark rate (about 250 Hz) and correlated pulses (30\%). 
\end{abstract}


\maketitle
\section{Introduction}

Several experiments have been conducted worldwide, with the goal of observing low-energy nuclear recoils induced by WIMPs scattering off target nuclei in ultra-sensitive, low-background detectors.  In the last few decades noble liquid detectors designed to search for dark matter in the form of WIMPs have been extremely successful in improving their sensitivities and setting the best limits. Current dark matter detectors using noble liquids have an effective target mass ranging from 100 kg to the ton-scale (e.g. LUX~\cite{[LUX]}, Xenon-1T ~\cite{[Xenon]}, DarkSide ~\cite{[DS50],[DS50a],[DS50b]}). Hundreds of 3 inch PMTs are used in these detectors for the accurate measurement of scintillation light from liquid argon (128 nm shifted to 420 nm) and liquid xenon (170 nm). 


An attractive alternative to photomultipliers is offered by silicon photomultipliers (SiPMs), a type of avalanche photodiode operated in Geiger mode, which have much lower intrinsic radioactive background~\cite{[Cebrian]} and smaller mass in addition to unrivalled performances in single photon detection. SiPMs behave linearly with a satisfactory gain of 10$^6$-10$^7$ and offer low intrinsic radioactive background, low operating voltage and power consumption, and have possibilities for inexpensive mass production.  
%
\subsection {Experimental setup and data acquisition}
\label{sec:data_taking}
\indent

\begin{figure}[!ht]
\center\includegraphics[height=7cm]{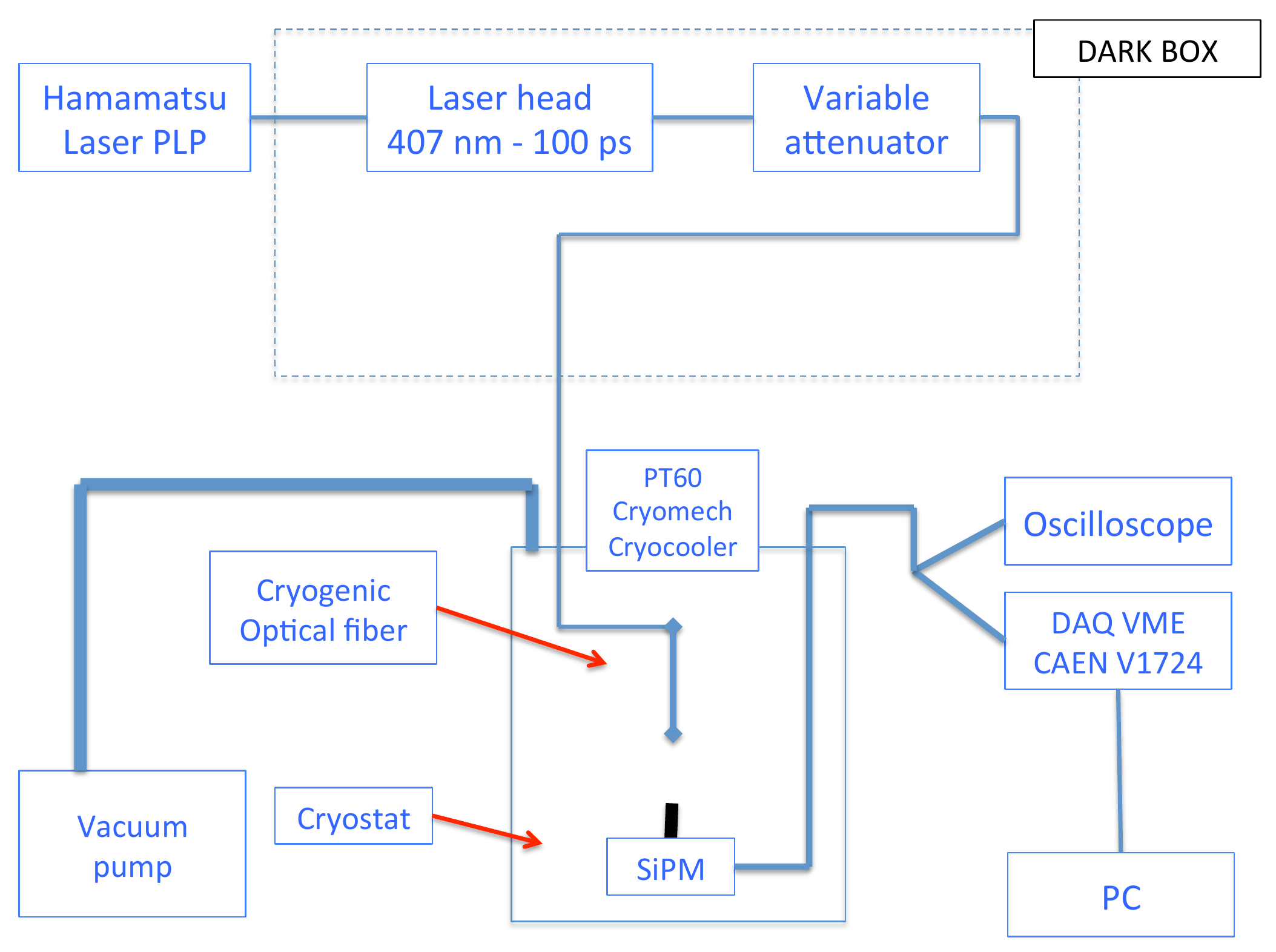}
\caption{Sketch of the experimental setup.} 
\label{fig:exp_sketch}
\end{figure}

In Figure~\ref{fig:exp_sketch} a schematic of the experimental setup is shown. The SiPM array, mounted on its cryogenic front-end board, is housed in a stainless-steel dewar of 25~cm diameter and 100~cm height, with an inner volume of approximately 50~liters. The dewar is closed by a stainless-steel flange equipped with a series of smaller size feedthrough flanges for: the liquid argon (LAr) input line, the evacuation line, the cryocooler head, the readout of the SiPM signals, supplying the bias voltage to the SiPM, connecting the temperature sensors, and for the transmission of laser pulses through the optical fiber.

\begin{figure}[!ht]
\center\includegraphics[height=7cm]{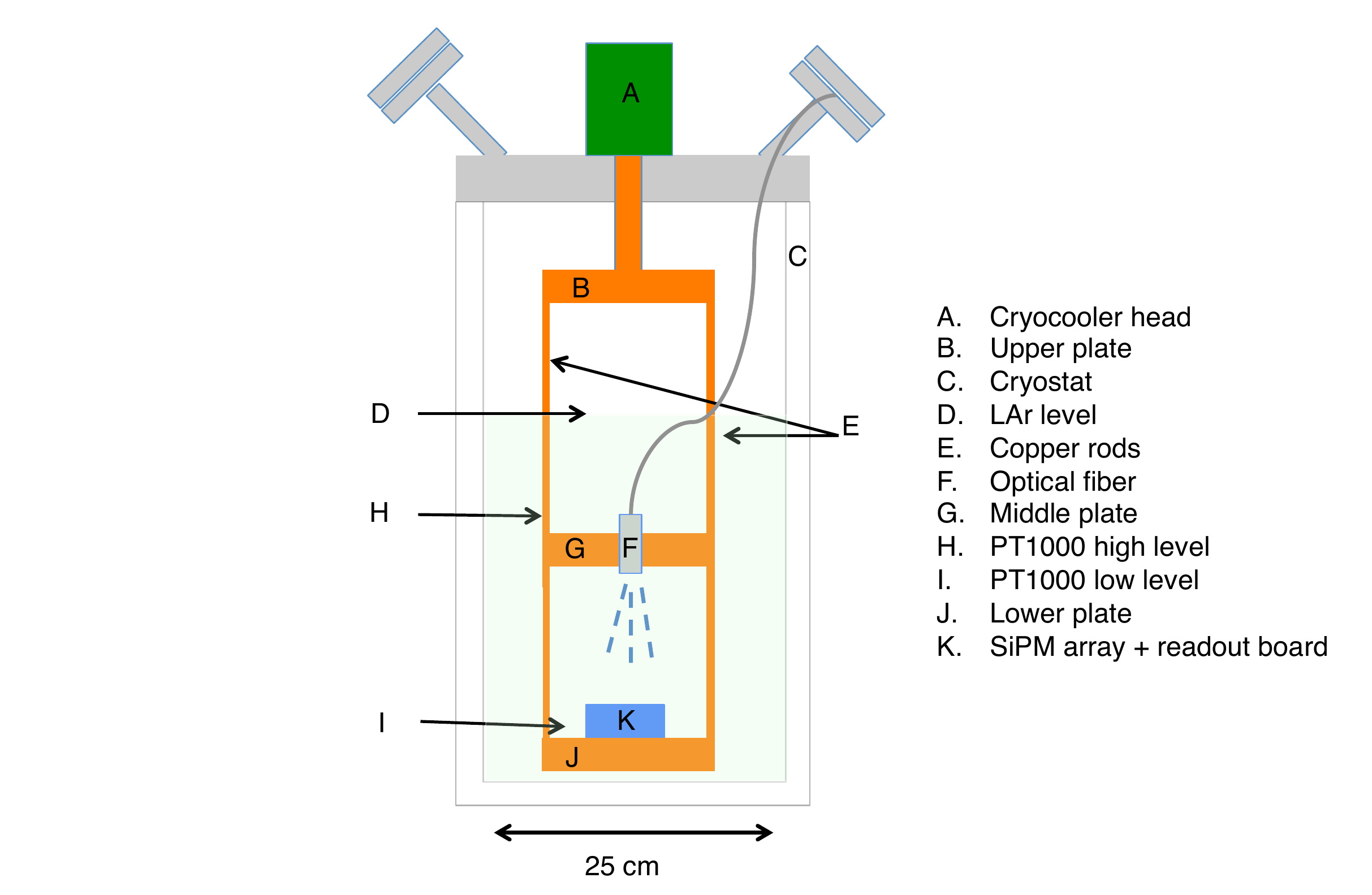}
\caption{Sketch  of the support structure of the photodetector: it consists of a set of three copper plates. The upper one is screwed to the cryocooler head for allowing a good thermal contact. The second and the third plate are placed at different heights  to host the optical fiber connector (middle plate) and the SiPM readout board (lower plate). Two PT1000 temperature sensors are placed on the middle and on the lower plate, acting as a level meter for the liquid argon.} 
\label{fig:copper_holders}
\end{figure}

The support structure of the SiPM array consists of a set of three copper plates (Figure~\ref{fig:copper_holders}). The upper one is screwed to the cryocooler head to allow for good thermal contact. In addition, three copper bars support to plates, placed at different heights to host the optical fiber connector (the middle plate) and the SiPM array readout board and a temperature sensor (lower plate). The plate hosting the optical fiber connector is fixed at about 10~cm above the SiPM array, so as to illuminate all the SiPMs dies. Light pulses are generated by a Hamamatsu PLP10 light pulser, equipped with a 408~nm laser head with a pulse width of 70~ps. The laser beam is passed through an optical attenuator mount in which discrete filters of various attenuation coefficients were used to select the magnitude of illumination reaching the photosensor.

Before the data taking starts, the cryostat is first pumped out until a residual pressure of 10$^{-4}$~mbar is achieved and then filled with LAr. The level of the liquid argon in the cryostat is monitored with two PT1000 resistors which are readout by a calibrated CRYOCON 32 controller. The first PT1000 (low level) is placed at the same height as the SiPM array, while the second (high level) is placed a few cm above the optical fibre output. The liquid argon filling operation is stopped 15~min after the high level is reached. The cryocooler maintains a constant LAr level, in order to ensure constant thermodynamic conditions throughout data taking. 

The Bias voltage was supplied via a low noise power\footnote{TTi QL355TP} supply through a 10~k$\Omega$ resistor. The SiPM array readout board, described in section~\ref{sec:readout}, has two outputs that are conveyed outside the cryostat through a signal feedthrough flange and are fed to an external (custom made) NIM amplifier (gain x10). The output of the amplifier is fed in to a CAEN V1720E digitizer with 4~ns sampling and 12~bit resolution, connected via optical link to a PC for data handling, storage and analysis.
%
%
%
\subsection {Cryogenic readout board}
\label{sec:readout}
\indent

Until very recently, the small size of the active area of SiPM dies were the main obstacle for them to be considered a valid alternative to PMTs in noble liquid direct dark matter search experiments. In the last few years much progress has been made to enlarge the effective area of silicon detectors by bonding SiPM dies together into arrays. These have now reached sizes as large as 5x5cm$^2$~\cite{[SensL]}. Large commercially produced arrays are usually provided with a common connector to bias all the dies and a number of output pins equivalent to the number of SiPM dies that are mounted in the array. In a large cryogenic apparatus {\em O}(10~ton) one should try to minimize the acquisition channels of the photosensors for several reasons. The number of signal output cables should be kept low because cables increase the heat load of the system and the radioactive budget. Moreover, a large number of channels increases both the complexity of the detector and the cryogenic system. It is therefore mandatory to find a solution for reducing the number of channels to be readout from a SiPM array. 

Our approach to solve this issue has been the use of an active front-end board placed near the SiPM array working at cryogenic temperature. The aim of the front-end amplifying board is to sum up the output of multiple SiPM dies without distorting the pulse shape (e.g. recharge time should stay constant). 

\begin{figure}[!ht]
\center\includegraphics[height=7cm]{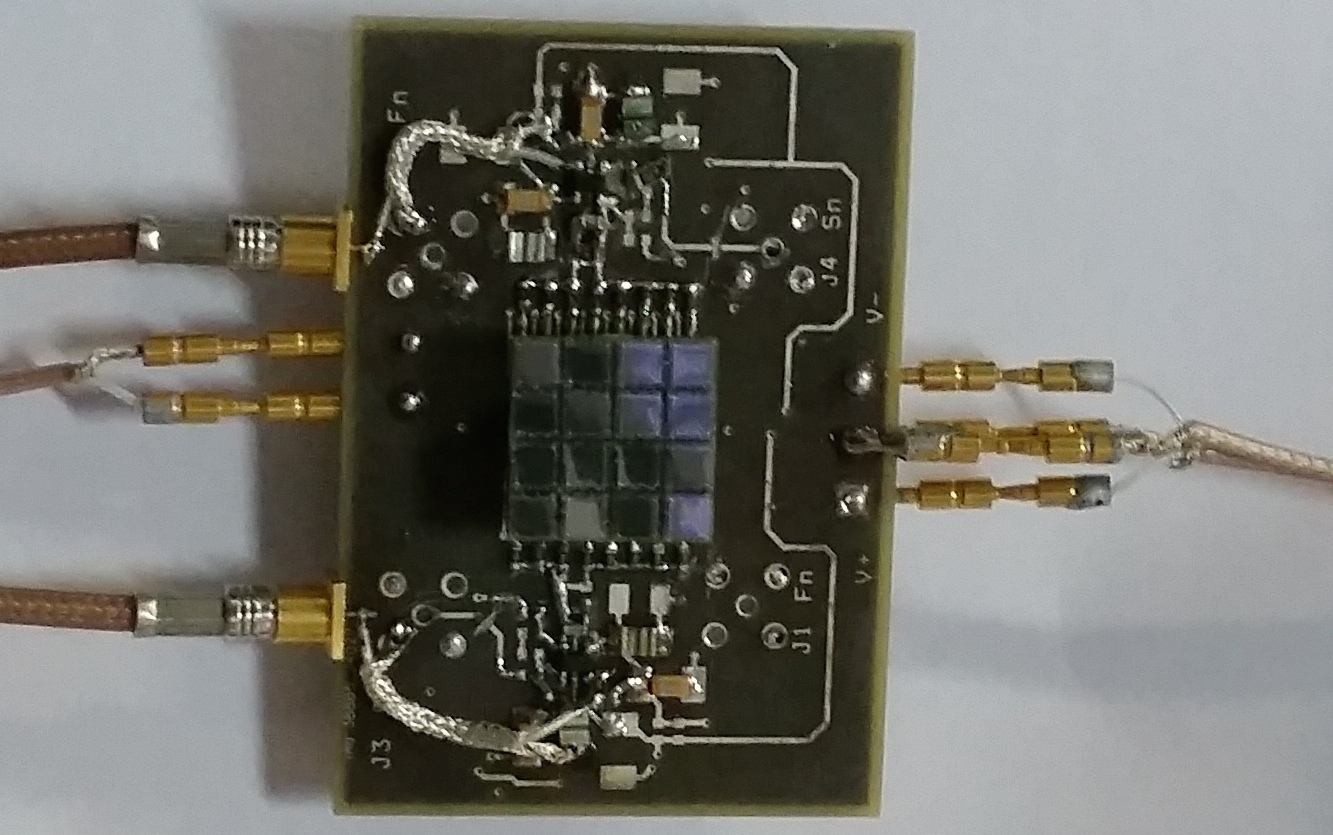}
\caption{Picture of the SiPM array and of the cryogenic front-end board.} 
\label{fig:array_pic}
\end{figure}

\begin{figure}[!ht]
\center\includegraphics[height=7cm]{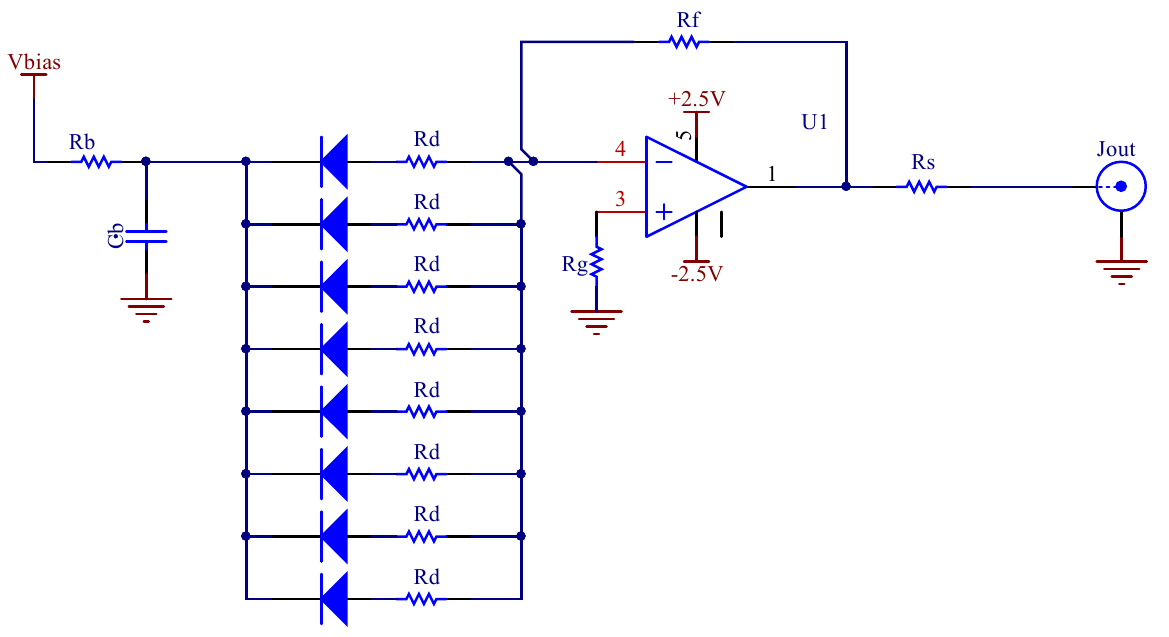}
\caption{SiPM readout board electric sketch. The feedback resistor value is R$_\mathrm{f}$=560~$\Omega$ and C$_\mathrm{b}$=220~nF.} 
\label{fig:electric_sketch}
\end{figure}

The front-end board is coupled to the SensL ArrayB-30035-16P (Figure~\ref{fig:array_pic}), which is composed of 4x4 SiPM dies of 3x3mm$^2$ each (see Table~\ref{tab:data_sheet}), and it conveys the output of 8 SiPMs dies together through the electric scheme shown in Figure~\ref{fig:electric_sketch}, resulting in two summed output channels of 8 SiPM dies each.

The board and the SiPM array showed good performance in terms of mechanical robustness, undergoing multiple cooling/warming cycles without any failure.  
\begin{table}[h!]
\begin{center}
\begin{tabular}{c c c c c c c}
\hline \hline
Model          &   Size    & PDE$_{max}$  & $\mu$cell size & cells   & Recharge time~[ns] & C$_\mathrm{SiPM}$ [pF]   \\
                &  mm$^2$   &    \%       &  $\mu$m        &         &      at 300~K     &   at 300~K   \\
    SensL MicroSB-30035  &   3x3     &    41       &      35        & 4774    &       180            &        850      \\
\hline \hline
\end{tabular}
\caption{Summary of the characteristics at room temperature (from data sheet) of the SIPM dies forming the array under test.}
\label{tab:data_sheet}
\end{center}
\end{table}
%
\subsection {Data Taking}
\indent

A study of the performance of the readout board coupled with the SiPM array has been performed at the liquid argon temperature. The main characteristics under study have been the breakdown voltage, the recharge time, the single photoelectron (SPE) spectrum and resolution,  the dark rate, the correlated pulses and the relative PDE. A scan of the performance of the SiPM array as a function of V$_\mathrm{bias}$ has been performed. Each data taking run consisted of 100.000 triggers at a given V$_\mathrm{bias}$, with a memory buffer of 5~$\mu$s, with 2~$\mu$s pre-trigger. For the SPE runs the laser was triggered by an external pulser at a repetition rate of 10~kHz and its illumination was set, through a system of discrete attenuators in order to send a few photons for each trigger to the photosensor.

The single photoelectron spectrum, was reconstructed by integrating the waveform region around the trigger for 800~ns\footnote{The integration window was set as large as to contain at least three time the recharge time (3$\tau$).} for a given run. 
%
%
%
\begin{figure}[!ht]
\center\includegraphics[height=7cm]{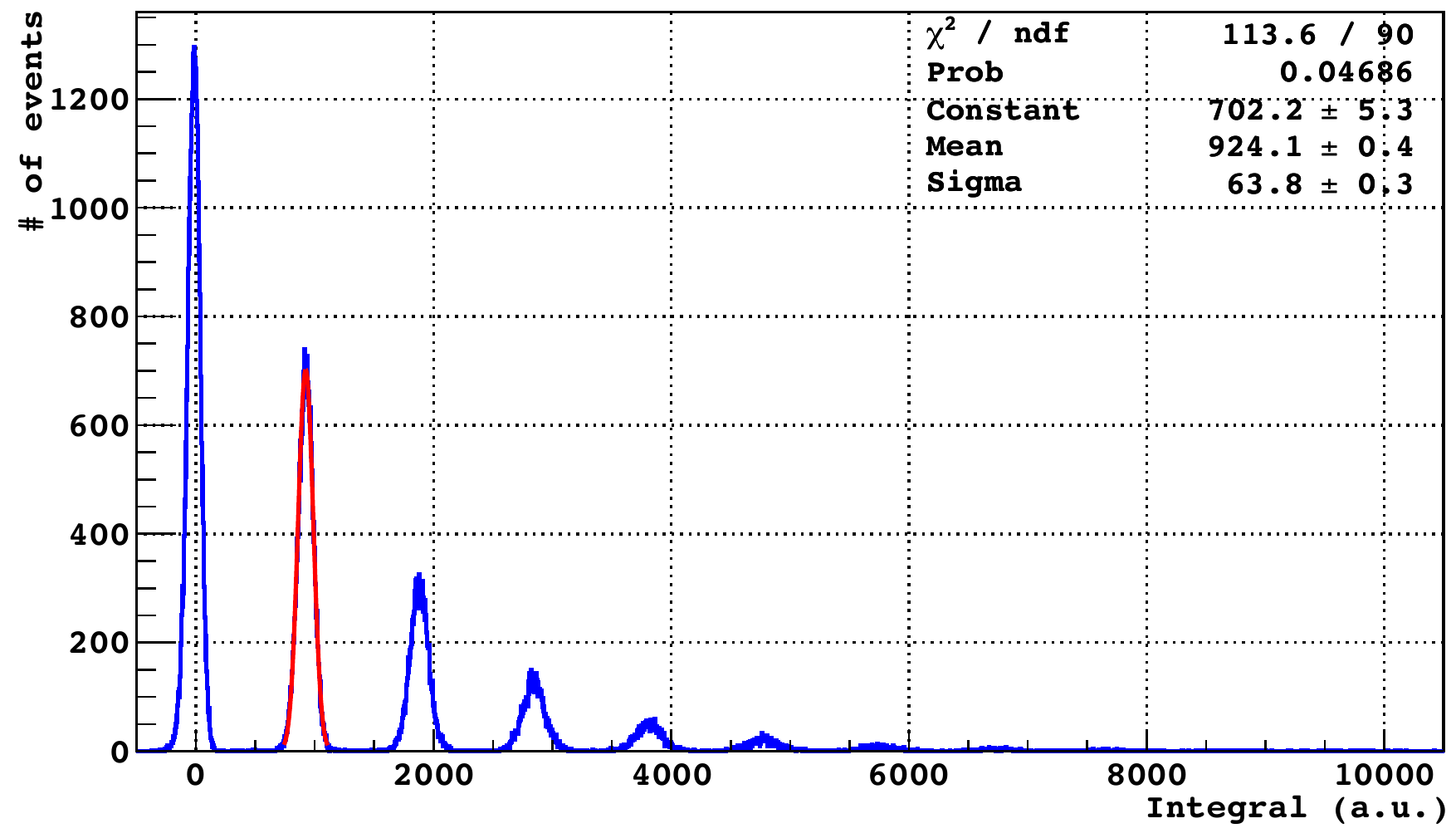}
\caption{Example of Gaussian fit of the first photoelectron spectrum.} 
\label{fig:SPE_fit}
\end{figure}
The SPE response of the SiPM array at each V$_\mathrm{bias}$ is calculated by fitting the first photoelectron peak of the SPE spectrum with a Gaussian shape, as shown in Figure~\ref{fig:SPE_fit}.
In addition, we define the SPE resolution as
\begin{equation}
Res=\frac{\sigma_\mathrm{SPE}}{\mu_\mathrm{SPE}}
\end{equation}
where $\sigma_\mathrm{SPE}$ is the sigma of the gaussian fit of the first photoelectron peak.
%
\section {Results}
\label{sec:analysis}
\indent
\subsection {SPE response and resolution}
\begin{figure}[!ht]
\center\includegraphics[height=7cm]{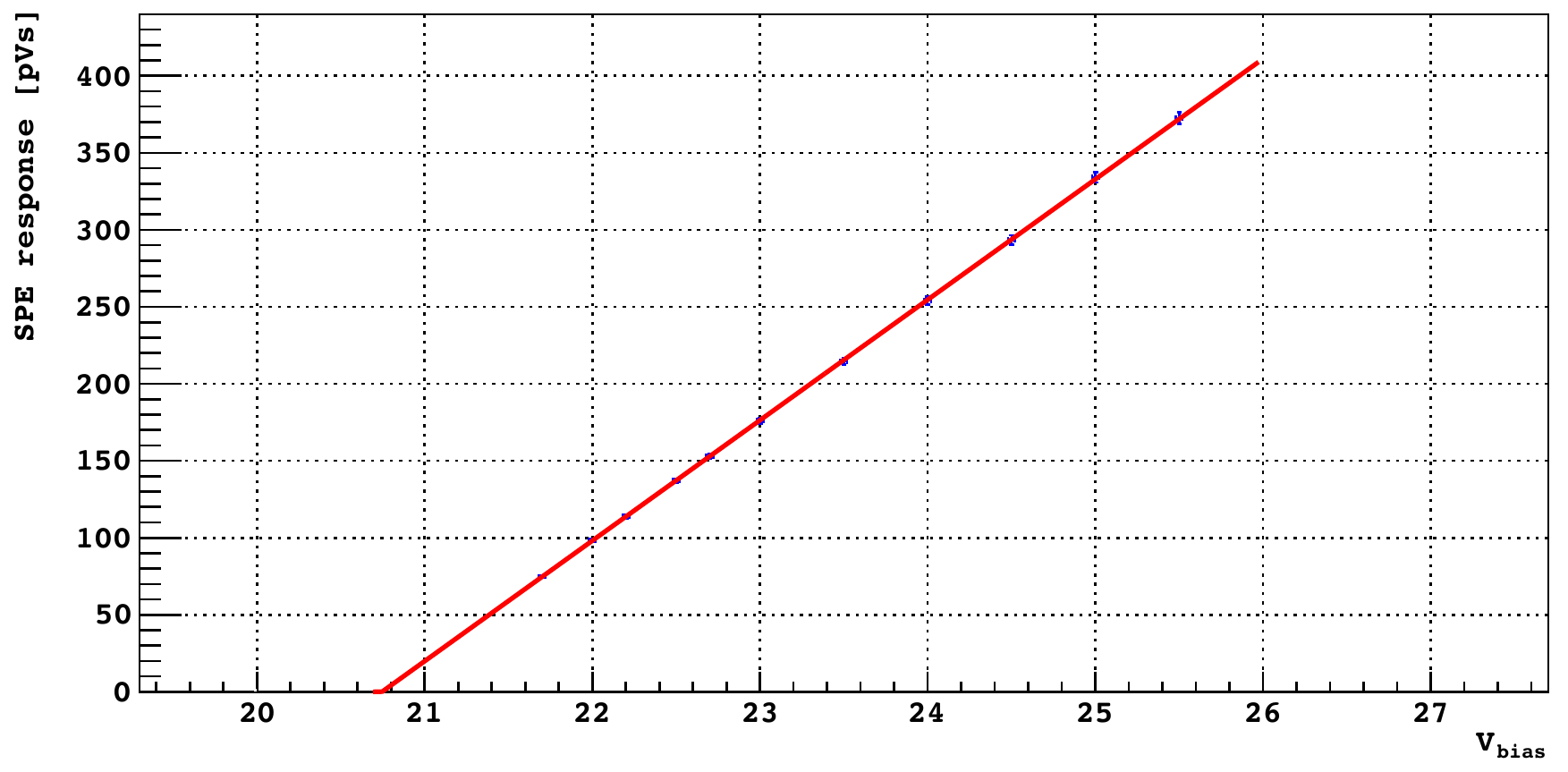}
\caption{SPE response of the SiPM array as a function of V$_\mathrm{bias}$. Points are fit with a line. Results of the fit indicates that $V_\mathrm{bd}^{87K}$=20.75$\pm$0.13~V.} 
\label{fig:workfunction}
\end{figure}

\indent
Figure~\ref{fig:workfunction} shows the SPE response of the SiPM readout board (workfunction) as a function of the V$_\mathrm{bias}$ at the liquid argon temperature. Points have been fit with line. The workfunction is found to be linear in the whole range of V$_\mathrm{bias}$  explored. The result of the fit indicates, as expected, that the $V_\mathrm{bd}^\mathrm{87K}$=20.75$\pm$0.13~V is much lower than the corresponding one at room temperature ($V_\mathrm{bd}^\mathrm{300K}\sim$25~V). 

An important parameter that stresses the great performance of the SiPMs with respect to the PMTs is the excellent SPE resolution. SiPMs can be as good as 4\% while cryogenic PMTs usually show performance of 25-35\%. Summing up several channels of a SiPM array through a readout board could spoil the SPE resolution for two reasons: firstly, there is a spread of $V_\mathrm{bd}$ among the SiPM dies installed on an array, and hence a slightly different gain for each channel, secondly, the large input capacitance of each SiPM die (about 1~nF) might increase the electronic noise and reduce the signal-to-noise ratio. 

\begin{figure}[!ht]
\center\includegraphics[height=7cm]{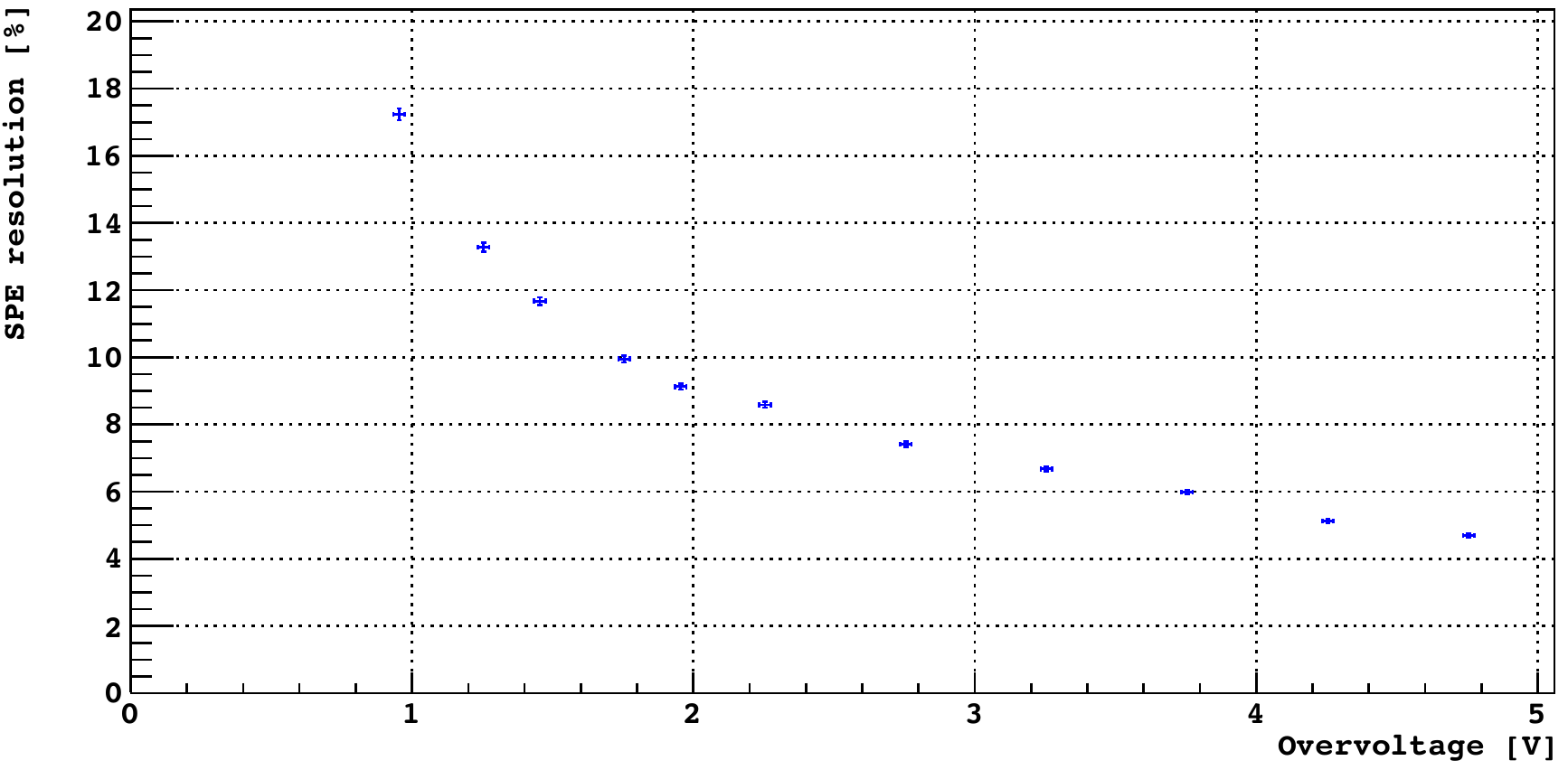}
\caption{Single photoelectron resolution as a function of the overvoltage.}
\label{fig:SPE-res}
\end{figure}

Figure~\ref{fig:SPE-res} shows the SPE resolution as a function of the overvoltage (V$_\mathrm{OV}$=V$_\mathrm{bias}$ -V$_\mathrm{bd}$). The SPE resolution decreases with the overvoltage as expected and it goes as low as 5\% at V$_\mathrm{OV}\sim$5~V. Results of our measurements are quite promising and highlight the encouraging performance of the readout board. At this stage, the possibility of summing up even larger-size arrays with more channels with the same technique seems possible. 
\subsection {Pulse shape studies}
\indent

Recharge time is an important parameter for a SiPM affecting the pulse shape. It is given by $\tau_{rec}=C_d\cdot R_q$ where $C_d$ is the diode capacitance and $R_q$ is the quenching resistor. For polysilicon quenching resistors,  $R_q$ increases with decreasing the temperature. Thus, the pulse duration at liquid argon temperature is larger than the corresponding one at room temperature. This feature might have an impact in liquid argon-based dark matter experiments since it might spoil the pulse shape discrimination capabilities in case the recharge time is too large with respect to the fast component of the scintillation light. A common integration window is 90~ns~\cite{[DS50b]}.
%
%
%
%

\begin{figure}[!ht]
\center\includegraphics[height=7cm]{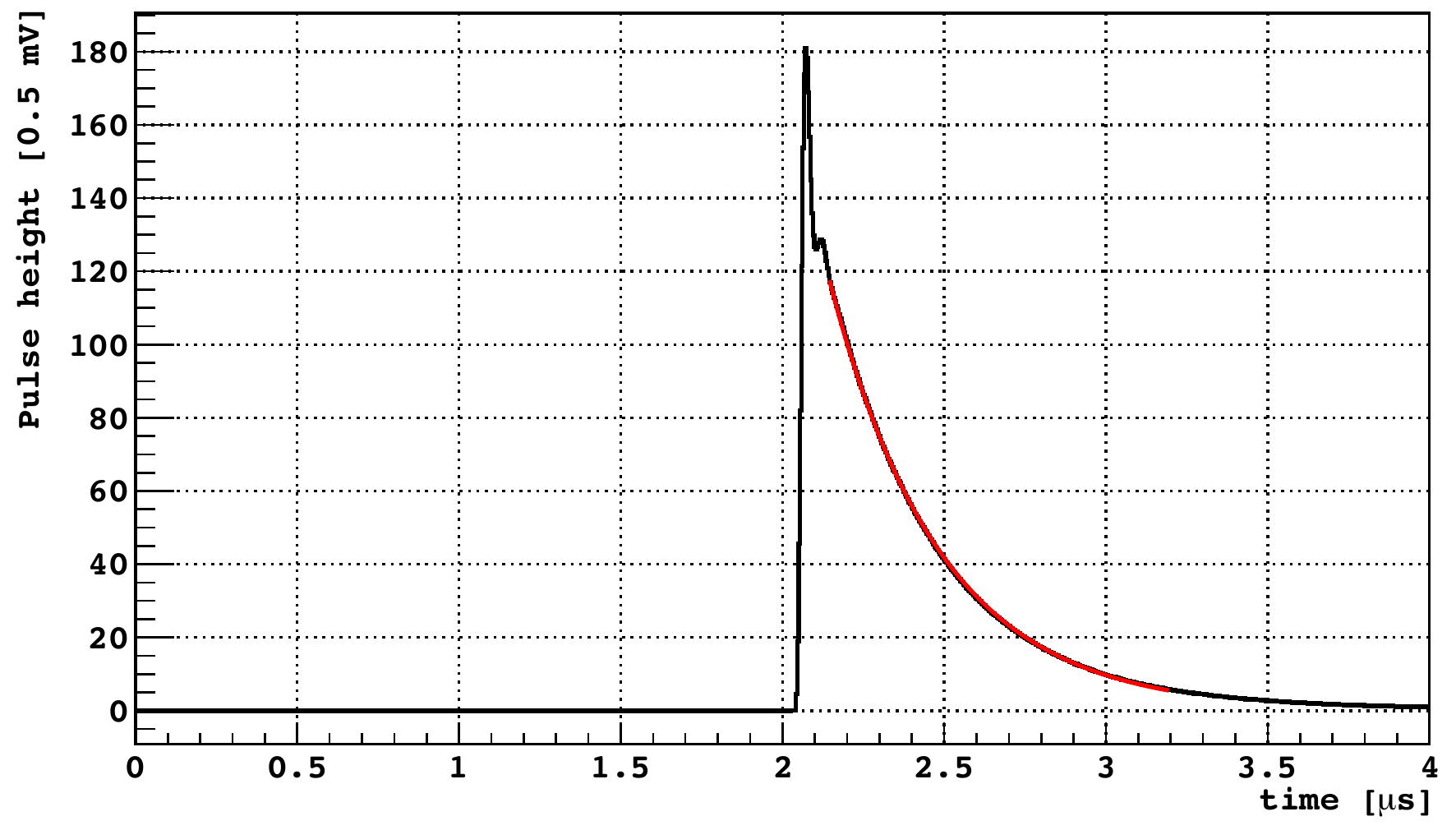}
\caption{Fit of the falling edge of the average (inverted) waveform of 100000 triggers at the liquid argon temperature.}
\label{fig:RecTime}
\end{figure}

We performed a precise evaluation of the recharge time constant of the 8 channel summed output by fitting the falling edge of the average waveform of 100k events with an exponential curve. Results of the fit are shown in Figure~\ref{fig:RecTime}, where the recharge time can be estimated to be $\tau=343\pm5$~ns. This results is consistent with the value we obtained by measuring the recharge time of one of the single SiPM dies forming the array at the liquid argon temperature.

\begin{figure}[!ht]
\center\includegraphics[height=7cm]{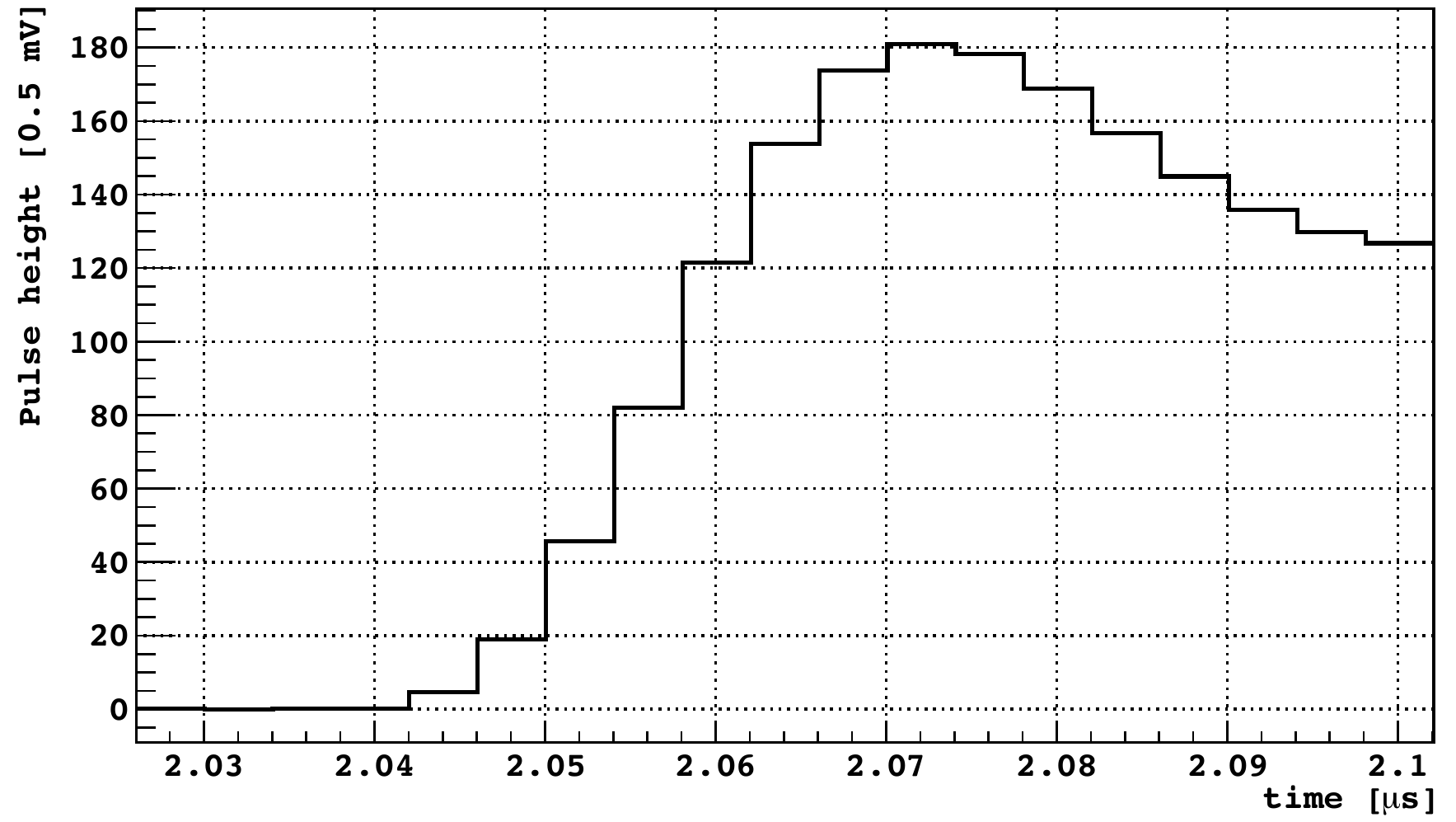}
\caption{Zoom of the the rise time region of the averaged waveform at the liquid argon temperature.}
\label{fig:Riseup}
\end{figure}

The parallel of the capacitance of the 8 summed SiPM dies reduces the bandwidth of the amplifier. In Figure~\ref{fig:Riseup}, a close-up of the rising edge of the average waveform of 100k events is shown. The resulting rising edge is  slowed slightly (about $\sim$20~ns) with respect to the rise time of a single SiPM die (a few ns). 
%
\subsection {Relative PDE and correlated pulses}
\indent

Every specific application of SiPM requires optimal selection of operating voltage to balance the signal performance (gain, photon detection efficiency, timing resolution, etc.) with noise (dark count rate, cross-talk, and afterpulsing).
This section reports about the measurements performed to highlight the influence of cross-talk and afterpulsing on photodetection characteristics that are useful for such optimization by using the method described in~\cite{[Vinogradov]}.

The typical experimental result of a few photon short pulse detection is the single photoelectron spectrum (SPE). Figure~\ref{fig:SPE_fit} shows an example of a histogram of SPE where the distribution of output charge reflects the probabilities to detect photoresponse pulses equal to 0, 1, 2, and more fired pixels (referred to as photoelectrons). The histogram peaks are very narrow due to low excess noise of charge multiplication, and thus the superb photon counting capabilities of SiPMs. In the absence of correlated pulses, the probability distribution should follow a Poisson law. However, when cross-talk and afterpulsing are considerable, deviations from Poissonian behaviour can be observed. 

We performed measurements of the relative PDE ($\lambda$) and of the correlated pulses as a function of $V_\mathrm{bias}$. Measurements have been performed through the following steps. The SPE spectra at each voltage and at different temperatures have been discretized, and the mean value and variance have been evaluated. Following~\cite{[Vinogradov]} the mean value ($\mu_{SPE}$) of the discrete SPE distribution can be rewritten as:
\begin{equation}
\mu_{SPE}=\frac{\lambda}{1-p}
\end{equation}
while the variance can be rewritten as:
\begin{equation}
Var=\frac{\lambda(1+p)}{(1-p)^2}
\end{equation}

The parameter $p$, the duplication probability, takes into account the deviations of the SPE spectrum from the Poisson law, while the $\lambda$ value is widely used in the evaluation of photon detection efficiency, when the mean number of photons per detected pulse $N_{ph}$ is known $\lambda=N_{ph}\cdot PDE$.

\begin{figure}[!ht]
\center\includegraphics[height=7cm]{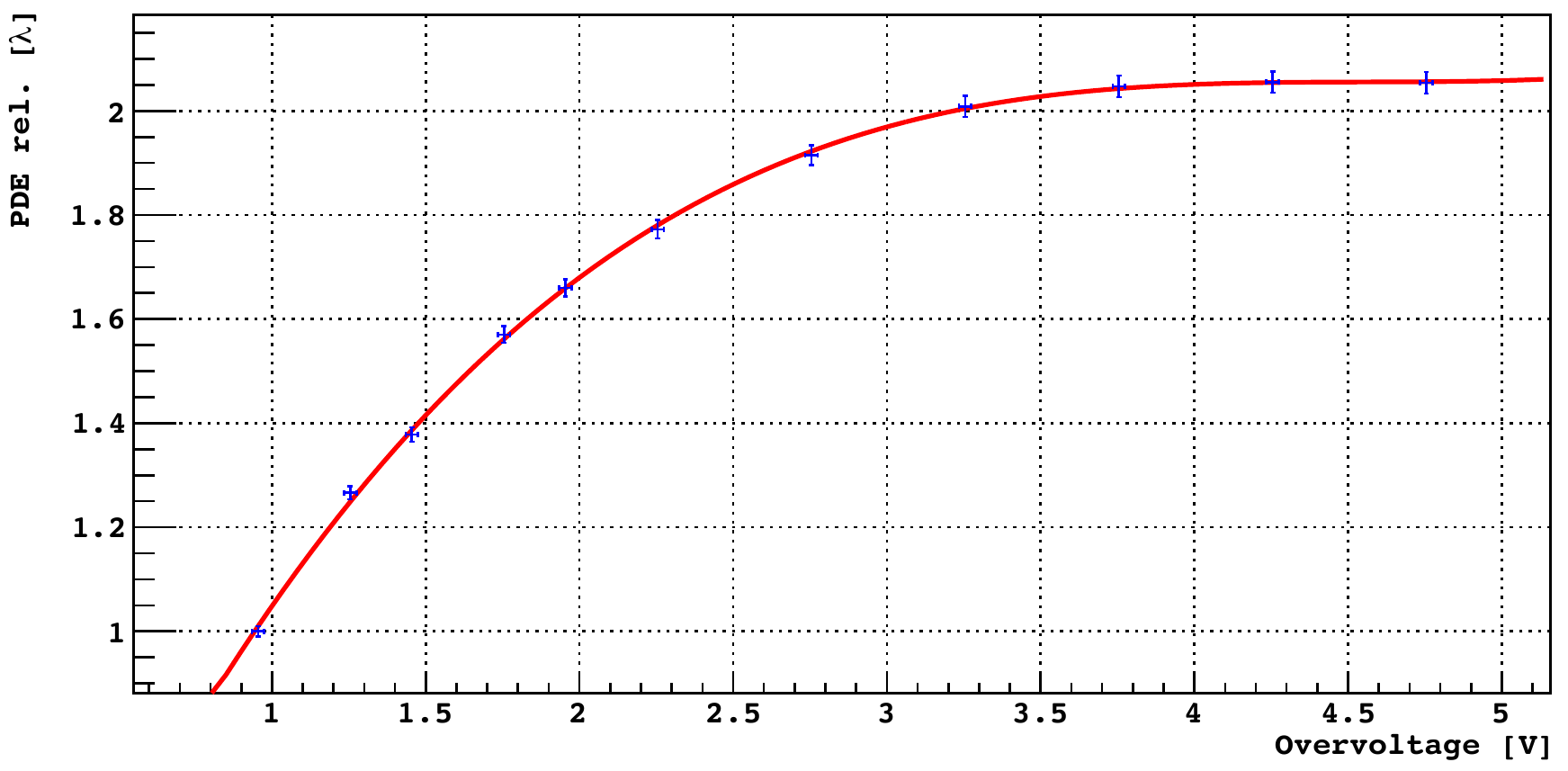}
\caption{Relative Photon Detection Efficency (PDE) as a function of the overvoltage.}
\label{fig:relPDEvsBias}
\end{figure}

Absolute PDE evaluation was not possible in our experimental setup. However, we performed a comparative measurement of the PDE as a function of the overvoltage, as shown in Figure~\ref{fig:relPDEvsBias}. Specifically, we plot the ratio of $\lambda$ at a given overvoltage divided by the $\lambda$ at 0.8~V overvoltage (at the smallest overvoltage value acquired). This plot shows that the PDE increases with overvoltage until it plateaus. It is worthwhile note that the maximal PDE is already reached at about V$_\mathrm{OV}$=3.5~V.

\begin{figure}[!ht]
\center\includegraphics[height=7cm]{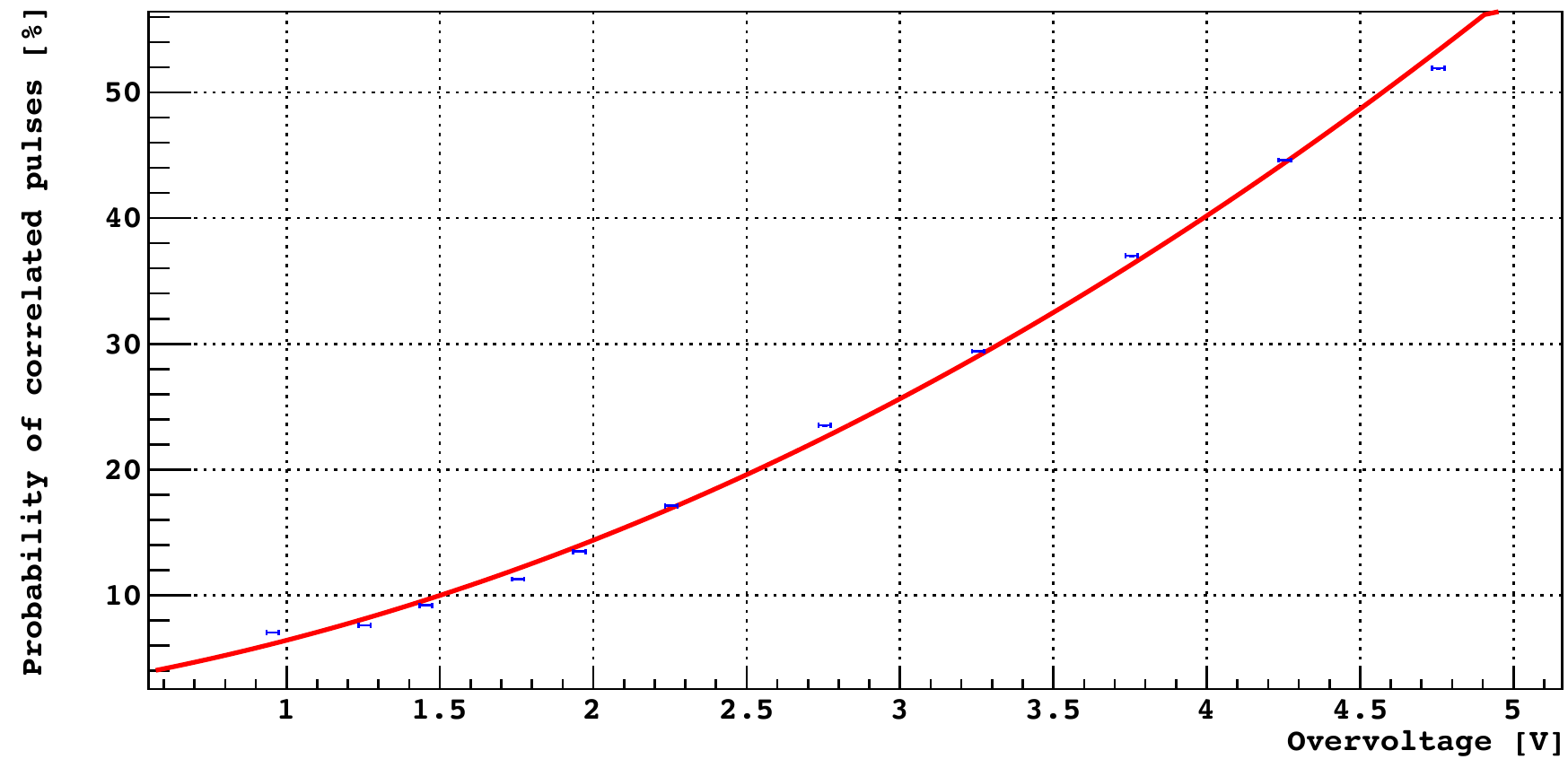}
\caption{Probability of Correlated pulses as a function of the overvoltage.}
\label{fig:pvalue}
\end{figure}

Once the V$_\mathrm{OV}$ has been set to maximize the PDE, the other figure of merit can be estimated. Figure~\ref{fig:pvalue} shows the probability of having a correlated pulse as a function of the overvoltage. The probability of correlated pulses increases like a pol2 function with the V$_\mathrm{OV}$ and at V$_\mathrm{OV}$=3.5~V it is about  30\%.
%
\subsection {Dark rate at liquid argon temperature}
\indent

\begin{figure}[!ht]
\center\includegraphics[height=7cm]{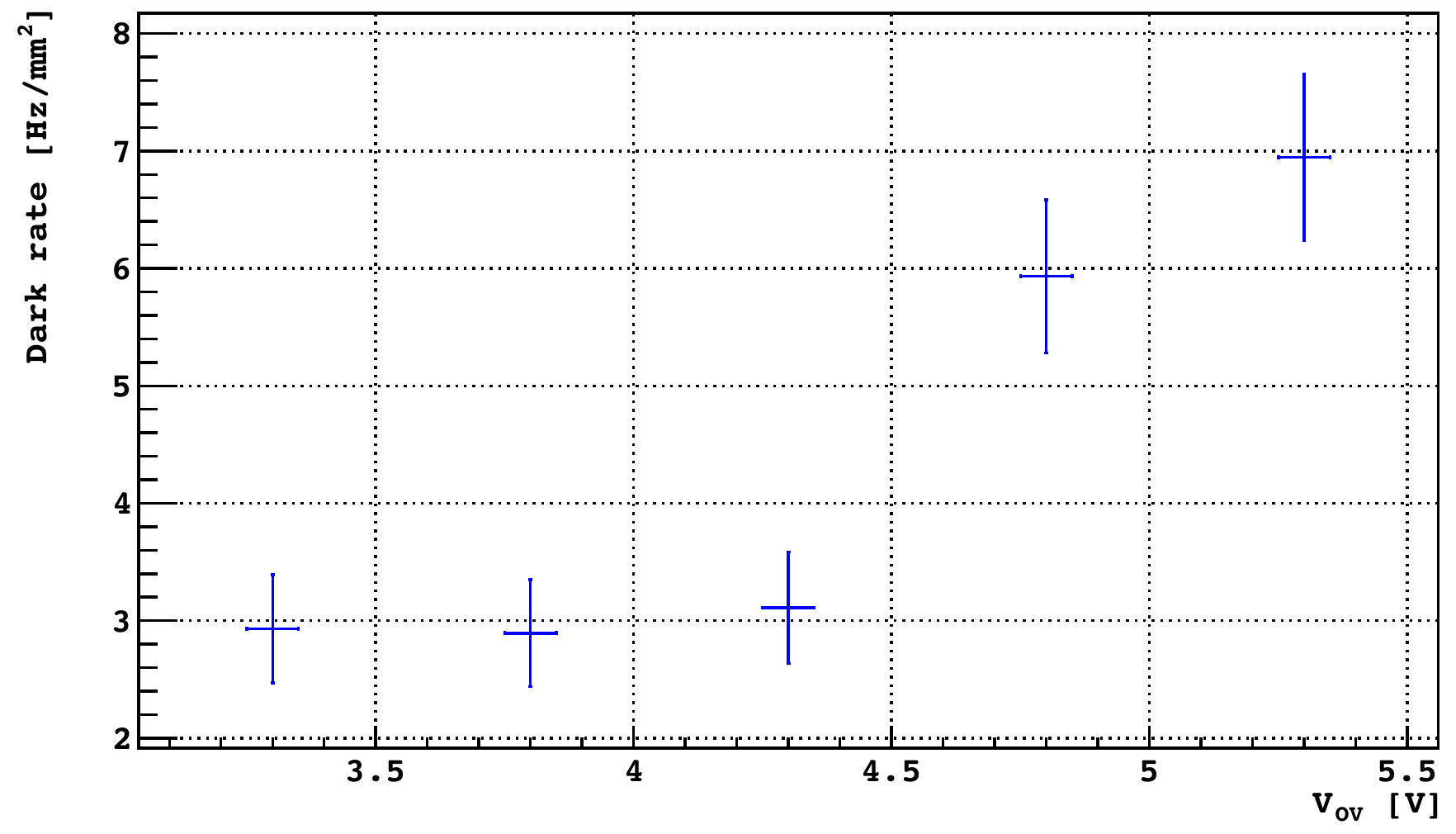}
\caption{Dark rate as a function of the overvoltage at the liquid argon temperature.}
\label{fig:dr}
\end{figure}

The dark rate has been evaluated through a peak finder algorithm that searched for peaks in the acquisition window of each triggered waveform for each run. The peaks found in the pre-trigger time window (i.e. before the reference laser pulse) have been counted. The dark rate is calculated with the following formula:
\begin{equation}
DR\mathrm{[Hz/mm^2]}=\frac{N_\mathrm{pulses}}{T_\mathrm{pre}~E_\mathrm{number}^\mathrm{tot}~A~N_{dies}}
\end{equation}
where $T_\mathrm{pre}$=2$\cdot10^{-6}$s is the pre-trigger window, $E_\mathrm{number}^\mathrm{tot}$=100000 is the number of triggered events for each run, $A=9~mm^2$ is the active surface area per SiPM die and $N_{dies}=8$ is the number of SiPM dies summed up by the readout board. The measured dark rate (expressed in Hz/mm$^2$) as a function of the overvoltage V$_\mathrm{OV}$=V$_\mathrm{bias}$-V$_\mathrm{bd}$ at the liquid argon temperature is shown in Figure~\ref{fig:dr}. This figure of merit surpassed the expectations, being four orders of magnitude smaller than the corresponding dark rate at room temperature ($\sim$MHz/mm$^2$). With this value a SiPM array equivalent to a 3 inch PMT (3000~mm$^2$) would have a dark rate of about 10~kHz at the liquid argon temperature.
This result is very promising and sets a very important milestone in the possibility of replacing cryogenic PMTs with large-size SiPMs arrays.  

%
\section {Conclusions}

SiPMs appear to be very promising devices for next generation noble liquids direct dark matter search experiments. SiPM arrays sizes are nowadays comparable to PMTs of 3 inch size. The manufacturing progress of last years have made SiPMs arrays very appealing for substituting PMTs in cryogenic environment, as also shown in~\cite{[Whitt]}. In particular, SensL-ArrayB-30035-16P can be operated at liquid argon temperature coupled with a cryogenic readout amplifying board, to reduce the number of output channels without distorting the pulse shape. Moreover, the array performance at V$_\mathrm{OV}$=3.5~V (corresponding to a gain of about 3$\cdot10^6$) where the PDE is maximal, are very promising in terms of SPE resolution (about 8\%), dark rate (about 250~Hz for the whole array) and correlated pulses (30\%).
%
%
%
\section* {Acknowledgments}
The work presented in this paper was conducted thanks to grant PHY-1314507 National Science Foundation and from INFN. We warmly acknowledge all the institutions. We thank A. Razeto for the useful support he provided and D. Sablone for the assembling of the cryogenic readout board. We thank L. Tatananni, N. Canci and the LNGS mechanical workshop for the realization of the cryostat top flange. We also acknowledge A. Anastasio, A. Boiano, P. Di Meo and A. Vanzanella and the electronic workshop of INFN Napoli for the fruitful discussions and their support during the assembling of the experimental setup. 

\end{document}